\documentstyle[12pt]{article}
\begin{document}
\hoffset 0.5cm
\voffset -0.4cm
\evensidemargin 0.0in  
\oddsidemargin 0.0in
\topmargin -0.0in
\textwidth 6.7in
\textheight 8.7in

\begin{titlepage}

\begin{flushright}
PUPT-1779\\
hep-th/9803191\\
March 1998
\end{flushright}

\vskip 0.2truecm

\begin{center}
{\large {\bf DLCQ-M(atrix) Description of String Theory, and Supergravity.}}
\end{center} 

\vskip 0.4cm

\begin{center}
{Gilad Lifschytz}
\vskip 0.2cm
{\it Department of Physics,
     Joseph Henry Laboratories,\\
     Princeton University, \\
     Princeton, NJ 08544, USA.\\ 
    e-mail: Gilad@puhep1.princeton.edu }

\end{center}

\vskip 0.8cm

\noindent {\bf Abstract}
We discuss the connection between Matrix string theory and the 
DLCQ of string theory.
Using this connection we describe the sense in which  perturbative string 
amplitudes are  reproduced in
the Matrix string theory.  Using recent realization of 
the connection between SYM and Supergravity,
we  suggest  how to describe   Matrix theory with  non-flat backgrounds.
\noindent                
\vskip 6 cm

\end{titlepage}
\section{Introduction}
Recently Seiberg \cite{seib} considered the DLCQ formulation of M-theory, 
and showed that it is equivalent to the BFSS formulation of M(atrix) theory
\cite{bfss}, as conjectured in \cite{sus}. 
The starting point is  M-theory (or string theory), compactified on a 
vanishing small circle.
M(atrix) theory (on $T^d$) is then seen to be the theory of D-d-branes 
in a certain
limit \cite{seib,sen}, in which the string length and the string coupling 
are taken to zero
but a certain combination of them stays fix.

What is the meaning of this limit ?.
The connection with DLCQ description  comes up as we define the sector
with $P_{-}=N/R$ of DLCQ  as a limit of
vanishing compact direction with $N$ units of momenta in this direction 
\cite{bbpt}.
This can also be seen by formally boosting a theory in a compact direction 
and taking the radius of
this direction to shrink thus defining a limiting procedure for null 
compactification.

More recently, using argument relating Black holes to D-branes, It was 
argued \cite{mald}, using \cite{ms,k1,gkt,gk}, that conformal SYM theory 
properties can be calculated in a 
region where perturbation theory is not valid, by supergravity on a certain
background. Similar argument for other SYM were given in \cite{imsy}.
 The precise identification for the conformal SYM was
given in \cite{gkp,wit4}. The limits taken in \cite{mald,imsy} are the same
as the ones taken for the DLCQ description.

One issue that needs to be addressed in DLCQ is the fate of 
states with $P_{-}=0$.
 One advantage of looking at DLCQ of string theory rather than M-theory is 
the issue of these zero-modes,
 which tend to make the loops diverge in field theory \cite{helpol}.
In the case of string theory this seems not to be the case \cite{bilal}.
 Then it is possible
to show that the SYM will reproduce perturbative tree level 
scattering amplitudes, in
the appropriate limit. 
If one starts with a  non-trivial backgrounds, 
as we will see, it can be incorporated 
into the SYM description
of the DLCQ of string theory, using the recent connection with supergravity.

In section (2) we describe the M(atrix) description of Type IIA string theory 
\cite{motl,bs,dvv} following  Seiberg and Sen. 
We review  the  the degrees of freedom and 
the generalization to include certain backgrounds.
In section (3) we discuss scattering amplitude, and give a precise 
correspondence which shows the emergence of perturbative amplitudes in the 
M(atrix) string, we then discuss the supergravity 
side and show that the existence of 
non-renormalisation theorems needed for the M(atrix) theory to work are
compatible with \cite{mald,imsy}. In section (4) 
following \cite{gkp,wit4} we describe
how the states with no momenta in the compact vanishing direction couple to the
SYM. This gives a way to describe the effect of $P_{-}=0$ states, in particular
the effect of non-flat  backgrounds.

\section{Type IIA string Theory}

We start with Type IIA string theory with string coupling $g_s$ , 
 a  string 
length $l_s$, and a generic length L. 
We now compactify it on a circle of radius $R_s$ (which will be taken to zero) 
and put N 
units of momenta along
the circle. We will
use a set of duality transformation to map it into another string theory.
We first perform a T-duality along the circle this gives\footnote{Through out
the paper we drop factors of $2\pi$}
\begin{eqnarray}
g_{s}^{'} & = & g_{s}\frac{l_s}{R_s} \\ 
R_{s}^{'} & = & \frac{l_{s}^{2}}{R_s} \nonumber \\
L^{'} & = & L \nonumber
\end{eqnarray}
The N units of momenta along the circle transform to N Type
 IIB winding strings.
As the string coupling will become large we  perform a S-duality. This gives

\begin{eqnarray}
g_{s}^{''} & = & \frac{R_s}{g_s l_{s}^{2}} l_{s} \label{parast}\\
R_{s}^{''} & = & \frac{l_{s}^{2}}{R_s} \nonumber\\
L^{''} & = & L \nonumber\\
(l_{s}^{''})^2 & = & \frac{g_s l_{s}^{2}}{R_s} l_s.\nonumber
\end{eqnarray}
The string winding become D-string winding.
Already we see that any generic length $L$ is much smaller than the string 
length.
To get the SYM limit from the parameters above, we have
to  define a new scale. 
The SYM energies from  the parameters in equation(\ref{parast}) 
are all proportional
to $R_s$. One then just rescale  all energy scale by a factor $\bar{R}/R_s$ 
where $\bar{R}$ can be anything. If however we choose $\bar{R}=R$ then the
SYM description would match to the original starting point 
(DLCQ of Type IIA with parameter  $R$), with no other scaling necessary.
This means that all length scales change by $R_s/R$, including time.
The parameters of the 
Type IIB theory are now
\begin{eqnarray}
g_{s}^{n} & = & \frac{R_s}{g_s l_{s}^{2}} l_{s}\label{paranew} \\
R_{s}^{n} & = & \frac{l_{s}^{2}}{R} \nonumber \\
L^{n} & = & L R_s/R \nonumber \\
(l_{s}^{n})^2 & = & R_s\frac{g_s l_{s}^{2}}{R^2} l_s.\nonumber
\end{eqnarray}
and one arrives at the SYM theory.
The Yang-Mills coupling constant on the D-strings is
\begin{equation}
g_{ym}^{2}=g_{s}^{n} (l_{s}^{n})^{-2}=\frac{1}{g_{s}^{2}}
(\frac{R}{l_{s}^{2}})^2
\label{paraIIa}
\end{equation}
and the Higgs vev $L^{n}/(l_{s}^{n})^{2}$ and $vt/(l_{s}^{n})^{2}$ are finite.

In this limit the only surviving degrees of freedom are represented by 
SYM in (1+1) dimensions on a circle. 
The SYM Hamiltonian then represents the lightcone energy. It is given by
the total energy of the system minus the mass of the $N$ D-strings.

Equation (\ref{parast}) can be obtained also by starting with the original 
set up and instead of the T and S-dualities just exchange the coupling
 direction with the $R_s$ direction, using the M-theory interpretation.
However the first path can lead to formulations of backgrounds without a known 
M-theory interpretation as we will see below.

If we have originally $k$ compact directions $L_{i}$ we now T-dualise 
in directions $L_i$  to  get
\begin{eqnarray}
g_{s}^{''''} & = & \frac{R_s}{g_s l_{s}^{2}} l_{s}\prod_{i}^{k} 
\frac{l_{s}^{''''}}{L_{i}}=
(\frac{R_s}{g_s l_{s}^{2}})^{1-k/2} \prod_{i} \frac{l_{s}}{L_i} l_{s}^{1-k/2}\\
R_{s}^{''''} & = & \frac{l_{s}^{2}}{R_s} \\
L^{''''}_{i} & = & \frac{g_s l_{s}^{2}}{R_s}\frac{l_{s}}{L_{i}} \\
(l_{s}^{''''})^2 & = & \frac{g_s l_{s}^{2}}{R_s} l_s.
\end{eqnarray}
After scaling, the SYM coupling constant in ((1+k)+1) dimensions is then
\begin{equation}
g_{ym}^{2}=(\frac{R}{g_s l_{s}^{2}})^{2-k} \prod_{i} \frac{l_{s}}{L_i} 
\end{equation}

\subsection{Degrees of freedom}
Let us now look back and try to be more precise about the detailed
relationship between the Type IIA string theory and the 
SYM description\cite{motl,bs,dvv}.
The DLCQ description of the Type IIA is related under a Lorenz boost
to the Type IIA on a vanishing small circle. This under T, and S-dualities
is converted to a Type IIB theory of D-strings, as above. If we 
further scale the
parameters we can see that we end up with SYM.
On the Type IIA string 
theory side one has the massless modes (graviton, dilaton etc) and the
 massive modes of the string. On the Type IIB with D-strings  side the 
massless modes are 
constructed from the fermionic zero-modes of the D-strings. Winding states
around the vanishing circle are described by a state with momenta along
the direction the D-string is wrapped ($R_{s}^{''}$). The lowest unit
of momenta is $1/R_{s}^{''}$, This is true even if the string is multiply
 wound. However on the D-string one can have states with net momenta zero,
but energies $2/NR_{s}^{''}$ \cite{dasmat}, 
where $N$ is the wrapping number of the 
D-string. These are just states with open string running on both directions
of the wrapped D-string, with momenta $\pm 1/NR_{s}^{''}$. These represent
the massive states of the original string theory. This can be seen from
following the dualities and from the lightcone energies
\begin{equation}
E=\sqrt{(N/R_s)^{2} + l_{s}^{-2}}-N/R_s \sim  R_s /Nl_{s}^{2}.
\label{stex}
\end{equation}
From this one can see that in the SYM it cost less energy
to produce a massive state than a winding state in the large $N$ limit. 
In a  general scattering, these modes will be excited if there is enough  
energy as in
(\ref{stex}). From this and the uncertainty  relationship  
\begin{equation}
p^{max}_{\perp} \sim 1/b_{min}
\label{uncer}
\end{equation}
one  finds  $b_{min} \sim l_s$ which is expected from elementary string states.

The SYM description has more states than we have described. In particular it
has the degrees of freedom which are the off diagonal entries of the
matrices. What do these correspond in the original
string theory. At weak coupling string theory we do not see them, but the
SYM is T, and S-dual to a string theory on a small circle, for any string
coupling. Indeed looking at a string theory with a small circle, one can 
exchange the coupling direction with the small circle direction ($R_s$).
Now the new string length is $l_{s}^{''}$. There
could be excitations of open membranes 
where the wrapped membrane wraps around $R_s$ and
stretched between the two objects. This will have energy 
$\sim b/(l_{s}^{''})^{-2} \sim R_s$. Indeed these are the off diagonal
terms. In order for these to be excited in a collision of massless states, 
one needs 
 \begin{equation}
\frac{p_{\perp}^{2}}{2M} \sim \frac{b R_s}{g_s l_{s}^{3}}
\label{ener2}
\end{equation}
Where $M=N/R_s$  is the D-string mass.  Comparing equations
(\ref{stex}, \ref{ener2}) we see that as long as $b/R_{11}$ is large
the dominant excitations are the original massive string excitations 
and we are in the perturbative string regime. Once out of this regime we have
more possible excitations, and in the M-theory limit the new excitations
will dominate. When the stringy excitations are unexcitable,  there
is a new limit to the smallest distance one can probe,
$b_{min} \sim l_{p}$ \cite{dkps}.

\subsection{Type IIA Backgrounds}

It is also of interest to derive DLCQ  of Type IIA with some branes
present. After the sequence of T-duality and S-duality these branes will
become some other branes and the theory will then be modified from the SYM
in (1+1) dimensions to some other supersymmetric gauge theory that includes the
effect of the extra branes. Notice however that if we end up after the
dualities with a D-brane of dimension greater than one, the Yang-MIlls 
coupling constant on the brane will become zero in the limit 
$R_s \rightarrow 0$ 
, thus its own excitations do not contribute. In addition the brane can 
be boosted
 along their world volume or orthogonal to it. If they are boosted along 
they do not
pick up any momentum and eventually are wrapped on a circle of size $R_s$. 
In general
any other string state will pick up momenta and their
 interaction with the brane
will be represented in the gauge theory. This theory will have $8$ 
supercharges.
If we boost transverse to the brane it will pick up momenta that will be 
bound to it to form a BPS state. After the dualities the brane appears as a 
charge in SYM. In the following we give some examples.

\begin{itemize}
\item
start with Type IIA with D0 branes. After the set of 
dualities we end up with D1 branes and fundamental strings in direction $1$ The
theory on the D1 branes is then SYM in a sector with background electric
 charge \cite{dvv}. 
 
\item Start with $k$  NS-five brane with world volume directions 
$(0,1,2,3,4,5)$.
We will take $R_{s}$ to be always in direction $1$. After the T, and 
S-dualities this becomes a D-5 brane in directions $(0,1,2,3,4,5)$. The
D-1 branes are in direction $1$ with that direction having length
$R_{s}^{''}$ as given above.  The gauge theory Lagrangian includes 
other than the
Lagrangian of SYM in (1+1) dimensions $k$ extra hypermultiplets \cite{wit3}.

\item Start with the D4-brane in directions $(0,1,2,3,4)$. After the set
of dualities we get a D3-brane in directions $(0,2,3,4)$ and a D-1 brane in 
direction $1$. Again the length of direction $1$ is $R_{s}^{''}$.
If we are interested in the limit in which the D4-brane becomes an M-5
brane one should take the large $g_s$ limit. If one is interested in the
flat M-5 brane limit then we can T-dualise to get the theory of D4 branes
and D0 branes \cite{abkss}
. 
\item Start with the D6-brane in direction $(0,1,2,,3,4,5,6)$. After the
dualities we get a NS-five brane in direction $(0,2,3,4,5,6)$ and a 
D1-brane in direction $1$\cite{hanlif}.

\item Start with a D8-brane in direction $(0,1,2,3,4,5,6,7,8)$. This
does not have a known M-theory interpretation. This configuration is 
probably not a 
consistent type IIA configuration as the dilaton blows up a finite 
distance away from the
D8 brane. Never the less let us look at the transformation.  After the
T-duality we end up with a D7 brane and winding, after a S-duality we
get a (0,1) seven brane in direction $(0,2,3,4,5,6,7,8)$, and $N$  D1 branes
in direction $1$. If we had $k$ D8-branes we can end up with $K$ (0,1) 
seven branes.
The D1 branes can be outside the seven branes (that is separated in the 
$9$ direction), 
or the D1 brane can break between the (0,1) seven branes.
The theory on the D1 branes will then give the matrix formulation of the
 type IIA theory in a massive background. The details of the gauge theory
will be discussed elsewhere

\end{itemize}

\section{Scattering amplitudes}
In this section we will show that Seiberg prescription guarantees the correct
scattering amplitudes with no $P_{-}$ exchange, to be reproduced by the SYM.
When one starts with a Type IIA on a small circle, it might be that loops
become increasingly important as the circle shrinks.
 This however was shown not to be the case for the one loop \cite{bilal}, if
the on shell particles have momenta in the compact direction, and
it seems reasonable that this will hold to all loops\footnote{
The actual calculation in \cite{bilal} was for a one loop four point amplitude,
but the mechanism responsible for finiteness seem general}. 
This means that there
is a good perturbative expansion  for scattering amplitudes we are
interested in, of Type IIA on
a vanishing small circle. What we will see,  is that all scattering 
amplitudes have a common factor of $R_s$ which is the expected one from the
Lorenz transformation, and that after scaling, in the SYM it gives
 finite results. Using T and S-duality this means that tree amplitudes
are reproduced in the SYM. It is important to note that while 
the tree amplitudes with no $P_{-}$ exchange
are the same as those for ten-dimensional Type IIA theory, for finite $N$ ,
 the loop
correction while finite, are different then the ten-dimensional ones.

\subsection{Kinematics}
We consider scattering of ten dimensional massless particles.
The massless states have momenta
\begin{equation}
P^{i}=(\sqrt{(N_i/R_s)^2 + \vec{p}_{i}^{2}}, N_i/R_s, \vec{p}_{i}).
\end{equation}
All the kinematic variables $(P_i  + P_j)^2$ are finite, and so are the 
term involving 
polarizations, this of course is a consequence of Lorenz invariant.

In order to compute S-matrix elements one has to normalize the states 
appropriately. 
In order to compare later to the SYM we chose the non relativistic 
normalization for the states.
Each incoming and outgoing state gives a factor of $\frac{1}{\sqrt{2P_0}}$
As
one direction is compact with radius $R_s$ each of the incoming and 
outgoing states gives a 
factor of $\frac{1}{\sqrt{R_s}}$.  In addition the delta function of 
the momentum in
the compact direction becomes a $R_s \delta_{(\sum N_i, \sum N_f)}$. 
If The particles did not have any momenta in the compact direction 
then the $n$-point
amplitude
would have had an explicit factor of $R_{s}^{1-n/2}$ which would have given 
an effective 
string coupling $g_{s}/\sqrt{R_s}$ as one expects from low energy 
considerations.
 In our case
however $n$-point amplitudes just have a factor of $R_s$. Thus its 
Fourier transform
(the potential) will also have a $R_s$ factor, which is what 
happens in the T- S-dual 
description. 
In both case the reason is the same. The time has an explicit $1/R_s$ factor, 
picked up from
the Lorenz transformation.

These Kinematic factors multiply the usual string amplitudes as 
computed for instance in \cite{gsw}. The tree level amplitudes with no
momentum exchange in the compact direction, do not know if
there is a compact direction (other then through the normalization of 
states as discussed above), and are only a function of the Lorenz invariant
variables. Thus they give the same result as if we had ten large dimensions,
and this is true for any $N$.
In the loops the zero-modes run also and the loops are slightly different than
the ten-dimensional loops but are still finite \cite{bilal}. The difference
between the loop result and the ten dimensional loop result differs only
in an overall factor depending on $N_{i}$, this may be the origin
of the discrepancy found in \cite{bb,kp1}.
If we look
for long distance results (or at small $g_s$) 
the tree level will be the leading term, and thus
will be reproduced in the M(atrix) theory. It seems that in the large
$R$ and large $N$ limit, the loops will also match and one will recover 
lightcone string theory in ten dimensions. 

We would like to note that we do not actually know how to calculate these
scattering amplitudes in SYM, and the discrepancy with the tree
three-graviton scattering still needs to be solved \cite{dinerag}, 
but given theses arguments it
seem to be more of a technical problem than a real discrepancy.

\subsection{The Usual Example}
Let us look at the scattering of two massless states in the 
long distance limit. 
In the string theory for small momentum transfer one has
\begin{equation}
A_{4} \sim g_{s}^{2}\frac{R_s}{N_1 N_2} \frac{s^2}{t}
\end{equation}
where 
\begin{equation}
s=(P_1 +P_2)^2=\frac{1}{2}N_{1} N_{2}(\frac{\vec{p}_{1}}{N_1}- 
\frac{\vec{p}_{2}}{N_2})^2.
\end{equation}
If there is no change in $N$ of the scattered particles then
\begin{equation}
t \sim (\vec{p}_{1}-\vec{p}_{1}^{'})^2
\end{equation}
and one can easily define a non-relativistic like potential by Fourier
transformation. If the particles after scattering have a different $N$,
the amplitude is still well defined. In the large $N$ limit exchange
of a unit of momenta in the compact direction will result in a very
small change of $t$.

On the D-string side the potential between two moving D-strings
is given by\footnote{for Four graviton scattering in the orbifold
limit of SYM see \cite{af}}
\begin{equation}
V \sim N_1 N_2 \frac{R_{s}^{''} (\Delta v)^4 (l_{s}^{''})^4}{r^6}
\end{equation} 
using  $v= \frac{\vec{p}}{N/R_s}$, and equation (\ref{parast}),
 this matches the elementary string result. To get 
the DLCQ result, after the Lorentz transformation,  one
has to multiply by another factor of $R$.

Notice that what we call $v^4$ term, comes from a Lorenz
 invariant expression ($s^2$),
and as such is Lorenz invariant \cite{bbpt}.

\subsection{Supergravity Side}

We have started with a type IIA string theory on a small circle, with
$N$ units of momenta along direction $1$.
The metric associated with a wave with $N$ units of momenta in a compact
direction $R_s$ is given by
\begin{eqnarray}
ds^2 & = &  -dt^2(2-H)+dx_{1}^{2}H+2(H-1)dtdx_{1}+dx_{i}dx^{i}\\
H & = & 1+c\frac{g_{s}^{2} l_{s}^{8}N}{R_{s}^{2} r^6} \nonumber
\label{wave}
\end{eqnarray}
where $c$ is a numerical constant.
We would like to  calculate the scattering of  another particle, 
with momenta $n/R_s$ in direction $1$
and transverse momenta $\vec{p}$, from the above metric, 
in the supergravity approximation.
This can be done by evaluating the action of one particle in the 
presence of another, and
one can find a finite term even  in the limit $R_s \rightarrow 0$.

Instead of presenting the derivation in this fashion (see \cite{bgl})
 we would to do it in a way 
 to make contact with recent developments. Let us follow
the dualities,
using the rules for T and S-duality \cite{bho}, one gets
the solution for D-strings,
\begin{eqnarray}
ds^2 & = &  H^{-1/2}(-dt^2 + dx_{1}^{2}) + H^{1/2} dx_{i}dx^{i}\label{dst}\\
B_{01} & =  & -(H^{-1} -1)\nonumber\\
e^{-(\phi-\phi_{\infty})} & = & H^{-1/2}\nonumber
\end{eqnarray}
where $e^{\phi_{\infty}}=g_{s}^{''}$, and $x_1$ has periodicity 
$l_{s}^{2}/R_s$.
In the limit $R_s \rightarrow 0$
\begin{equation}
H \sim c\frac{Ng_{s}^{2} l_{s}^{8}}{R_{s}^{2} r^6} \nonumber
\end{equation}
and things again may look singular. This however  represent the
background fields that an original probe will feel, and
as scattering of the momenta states (in string theory)
was shown to be well define, one might expect a similar thing
to happen here (in the appropriate limit). To compute the long 
distance interaction, 
one couples the D-string to a background
generated by equation (\ref{dst}), using the Born Infeld action, plus
the WZW term.

A test D-string will couple through the action
\begin{equation}
S=T_{2}(\int d^{2} \sigma e^{-(\phi-\phi_{\infty})}\sqrt{\det{G}} +
\int B^{RR}).
\label{bi}
\end{equation}
where $T_{2}=\frac{1}{g_{s}^{''}l_{s}^{''}}=l_{s}^{-2}$, 
is finite as $R_s \rightarrow 0$.
We have suppressed here the dependence on the world volume gauge field of the 
probe, this can also be taken into account 
and will also produce finite results.

Taking a simple case where the test D-string is just rigidly 
moving with a velocity $v$, we evaluate the action (in static gauge) and find
\begin{equation}
S= T_{2} \int dt dx_{1} H^{-1} (\sqrt{1+Hv^2}-1)=\int dt[mv^2 +V(v,r)]
\label{pot}
\end{equation}
Now $H =c R_{s}^{-2}\frac{N g_{s}^{2} l_{s}^{6}}{r^6}$ 
but $v \sim R_s$ so inside the
square root every thing is finite. In addition $x_1$ has a periodicity of 
$1/R_s$ so after doing the $x_1$ integral the action $\sim R_s$. This
gives a potential which is $\sim R_s$ as we have found before. Now one can
scale all length scales by $R_s /R$ to get a finite potential.
This potential (from equation (\ref{pot})) 
coincides with the one conjectured to be the sum of planer 
diagrams in the SYM side \cite{bbpt,bgl,chtsy,kvpk},
 and indeed
checked up to two-loops.
It is interesting to note that a probe anti-brane, will also  produce a finite
result for a potential.

To make contact with \cite{imsy},
start with  equation (\ref{dst}) 
and scale all coordinates by a factor $R_s/R$. This gives the 
Type IIB the set of parameters 
of equation (\ref{paranew}), and in this limit we have seen all SYM
energies are finite. Now define $U=\frac{rR}{g_s l_{s}^{3}}$ and 
$\alpha = R_s \frac{g_s l_{s}^{2}}{R^2}=(l_{s}^{n})^2$. 
Then the metric becomes
\begin{equation}
ds^2 = \alpha [\frac{U^3}{g_{ym} \sqrt{cN}}(-dt^2 + dx_{1}^{2}) +
\frac{g_{ym} \sqrt{cN}}
{U^3}(dU^2 +U^2 d \Omega_7)].
\end{equation}
which is of course the same one given in \cite{imsy} 
(see also \cite{bps,hyun}). 
Notice that here however $x_1$ 
is compact with period $ l_{s}^{2}/R$, and that $t$ is not dilated any more. 

What is the range of validity of this metric ?.
The curvature of this solution is (in the original
Type IIA variables)
$\alpha^2 R^2 \sim \frac{r^2}{l_{s}^{2} N}$ while the local string coupling is
$e^{2\phi} \sim \frac{N l_{s}^{6}}{r^6} $. In addition because
 one has a compact direction 
it should be bigger than the string length giving the condition
$(\frac{r}{l_s})^2 \gg g_{s}^{4/3} N^{1/3}$. From this 
one might expect the calculation given in
equation (\ref{pot}) to be valid in the region (for small $g_s$)
\begin{equation}
N^{1/3} \ll  \frac{r^2}{l_{s}^{2}}  \ll N.
\label{sugralim}
\end{equation}
For large $g_s$ the left limit is replaced with $g_{s}^{4/3}N^{1/3}$.

On the other hand The SYM perturbative expansion, for this process, is valid if
\begin{equation}
\frac{r^2}{l_{s}^{2}} \gg N.
\end{equation}
Still particular terms in the SYM result seems to  agree 
with the supergravity
result.
 This was conjectured to be a consequence of non-renormalisation 
theorems on the
SYM side, which would have to be true for M(atrix) theory to work. 
Given that one gets
both the large $N$ SYM and supergravity limit ,as the same limit of 
parameters, 
one sees
that indeed the non-renormalisation theorem are correct (see \cite{bgl}
for another argument).

One can look at the opposite side,
why is it that the supergravity result seemed to be valid much beyond
it's naive validity. one possibility is as follows. 
Usually if the curvature becomes large then one can not trust the
classical solution, because higher curvature terms can change the solution. 
However the
background that we started with was that of a plane-fronted wave, 
and the potential derived
in equation (\ref{pot}) is the same as one could have derived from the plane 
fronted wave solution. This solution is known not to be corrected by
 higher curvature terms \cite{horsteif}.

\section{Non-trivial backgrounds}

We started off with Type IIA on vanishing small circle of radius $R_s$. 
The states that we have been considering are those with positive momenta in
 the compact direction. These are the states that in the DLCQ 
have non vanishing $P_{-}$.  
The massless states with $P_{1}=0$ (which define a non-flat background)
 transform under T and S-duality to be the massless elementary
string states in the Type IIB with D-strings, picture. One would like to know 
how to encompass them in the SYM description. 

As we have seen, the SYM that one ends up is  Type IIB with D-strings and
parameters given by equation(\ref{parast}). This limit is equivalent to
a limit in which both the string length and the string coupling constant
vanish (equation (\ref{paranew})). 
Still the metric generated by the D-strings although $\sim R_s$
gives finite results for  scattering of another D-string (as in the 
previous section).
As argued in \cite{mald} in this limit one gets on the
one hand SYM and on the other type IIB  on some space-time. The only 
difference between the case in \cite{mald,imsy}, is that the D-branes are
wrapped on a compact direction. Indeed the sigma model action
for a fundamental string on this background is  finite.

If the D-p-branes are not compact they result in a SYM (for $p>1$) with a 
moduli space. The propagation of
the Type IIB strings then gives the excitation in the particular 
vacuum defined by 
the moduli. If however the D-p-branes are compact then the SYM is
more complicated and in particular can move on it  classical vacuum manifold,
 which
is what is usually calculated in the M(atrix) theory. The energy
 dependence of a D-brane configuration is defined in the supergravity side
by equation (\ref{pot}), while on the SYM side it corresponds to calculation
of the vacuum energy. So there is an identification between vacuum energy and
the classical action of a probe in a supergravity background. Indeed the
SYM (or supergravity) gives the scattering off the DLCQ background, as one
wants in the M(atrix) theory.

The supergravity contains the effect of the D-branes on each other, as well
as the effect of perturbations of the background, on
the D-branes. Given that both
supergravity and SYM description are correct in a certain limit, it means
that in the M(atrix) model one can represent the
coupling to the background, as one expects in DLCQ. 

The guiding principle to describe the coupling are scattering calculations off
black-hole background and the corresponding SYM calculations in
 D-brane language.
This was exploited  in \cite{gkp,wit4} for the infinite D3-brane case.
These authors identified the SYM as the theory living on the boundary
 of $AdS_{5}$. Any perturbation on the AdS background induces a non trivial
function on the boundary. The correspondence between the SYM and the 
supergravity is then
\begin{equation}
W^{'}_{sym}(f(x)) =\ln Z_{string}(f(x,U))
\label{cores}
\end{equation}
The generator of connected Greens function 
$W'$ is evaluated using
the action
\begin{equation}
S_{ym}^{'}= S_{ym} + \int f(x) \Theta(x).
\label{action}
\end{equation}
and $f(x,U)$ are some perturbation around $AdS_{5}$, 
and $f(x)$ is their value
 at the boundary of $AdS_{5}$.
It is important to note that $f(x)$ has space like momenta \cite{gkp}, 
and thus is
a background\footnote{I would like to thank I. Klebanov and
O. Ganor for discussion on this point}.

For all other D-brane it is actually known that scattering
 calculations do not agree.
For the D1-brane they can agree in the correspondence 
point \cite{das}, and if one
uses the stretched throat they may scale correctly with energy,
 for all D-branes \cite{prm}.
The absorption cross section for gravitons  is computed from  
 the two-point function
of the energy-momentum tensor on the SYM side. The disagreement 
between the supergravity calculations and the SYM calculation 
can be taken as a sign of
large $N$ effects in the SYM \footnote{As the geometry is singular, 
for generic D-branes,
it is not clear how trustworthy the gravity calculations are. They 
can be a better 
approximation once
one looks at the thermal SYM in which case there is a large horizon 
on the gravity side. }

Let us view now the calculation of the previous section of scattering of
D-branes in this Language. While the above construction was only 
conjectured to be true for infinite D3-branes, let us for a moment extend it.
If for instance the D3-branes were compact, they could be moved with a finite
energy. One way to calculate the velocity dependent potential is to 
follow \cite{manton,ferear}.  Start with the 
solution describing static D3-branes. Then take one coordinate $r$ 
(moduli of the SYM) and
write everywhere $r=r+vt$. This is not a solution any more. 
Then one solves the equation of motion to some
order in $v$ and inserts to the classical action. 
From it one can read off the potential.
This is just another example of the general relationship, like equation
 (\ref{cores}). In the limit where the partition function is dominated by the 
classical minima the right hand side of equation (\ref{cores}) is just the
classical action as a function of some initial or boundary conditions.
The SYM gives those boundary (or initial) conditions for solving
the classical equations of motion, by specifying a fluctuation $f(x)$, and
 expectation values of fields (possibly time dependent). In general this
would then give a procedure for calculating scattering of D-branes in the
presence of some fluctuation, which is a manifestation of an
 originally non-trivial background. The problem however is 
that the supergravity solution is not well defined everywhere, 
and in particular for the D-string
is not valid at very large $r$. If there are no fluctuations,
instead of solving the classical equations
of motion which can be very hard,
we have already seen  there is a short cut to evaluate 
the velocity dependent potential.
One views  one of the groups of D-branes as 
a background and couple another one as a probe. This is just equation 
(\ref{pot}). For a certain region of $r$ this gives reliable answers.

Now what if we have a small perturbation above the background. The 
perturbation can be accounted for in the probe calculation by inserting the 
perturbed background. Here we are thinking of a perturbation above
the background of the target only, not taking into account the probe.
We now suggest what this correspond to in the SYM side.

In a similar fashion to \cite{gkp,wit4}, we suggest that
For each background field
(dilaton, graviton, etc) one adds to the SYM action a coupling
\begin{equation}
S_{ym}^{'}= S_{ym} + \int f(x) \Theta(x)
\label{action1}
\end{equation}
where $f(x)$ is the perturbation of  the background 
evaluated at the probe-brane
position,  
and $\Theta$ is 
an operator in the SYM.
For example, for the graviton $\Theta$ is the energy momentum
 tensor of the SYM. 
One can get the form of the operator (at least the bosonic part)
from the Born-Infeld action, alternatively one 
can look at scattering in M(atrix) theory and deduce, 
to leading order in energies,
The coupling of a generic state
to the massless modes.
In the case of  D0-brane this was described in \cite{kata}.

Calculation of the vacuum energy in SYM (properly Higgsed) should
now give the 
result corresponding to the perturbed background. This would hopefully resolve
the issues discussed in \cite{dos,do}.
Of course as the D-branes move the function $f$ will depend
on their position. In more complex situations 
one might need  to use a more exact treatment like
in \cite{doug,wpp}.

\pagebreak

\centerline{{\bf Acknowledgments}}
I would like to thank O. Bergman, O. Ganor, 
D. Kabat, I. Klebanov and  S.D. Mathur 
 for helpful  discussions.

\end{document}